\documentclass[reprint,aip,amssymb,jap, groupedaddress]{revtex4-1}
\usepackage{amsmath}
\usepackage{amssymb}
\usepackage{amsthm}
\usepackage{graphicx}

\begin{document}
\title{Controlling the quality factor of a tuning-fork resonance between 9 K and 300 K for scanning-probe microscopy}
\author{Georgios Ctistis}\email{g.ctistis@utwente.nl}
\author{Eric Frater}
\author{Simon R. Huisman}
\author{Jeroen P. Korterik}
\author{Jennifer L. Herek}
\author{Willem L. Vos}
\author{Pepijn W. H. Pinkse}\email{P.W.H.Pinkse@utwente.nl}
\affiliation{MESA+ Institute for
Nanotechnology, University of Twente, 7500 AE Enschede, The Netherlands}


\begin{abstract}
We study the dynamic response of a mechanical quartz tuning fork in the temperature range from $9\ \textrm{K}$ to $300\ \textrm{K}$. Since the quality factor $Q$ of the resonance strongly depends on temperature, we implement a procedure to control the quality factor of the resonance. We show that we are able to dynamically change the quality factor and keep it constant over the whole temperature range. This procedure is suitable for applications in scanning probe microscopy.
\end{abstract}

\maketitle
%

\section{Introduction}
Quartz crystal oscillators are widely used as a frequency reference. The mechanical resonance is
intrinsically stable and the crystal's piezo-electric effect allows for direct electronic interfacing.
Quartz tuning forks have proven to be extremely useful as sensors for temperature \cite{Smith1963, Jun2007}, mass \cite{Sauerbrey1959},
pressure \cite{Stemme1991, Kokubun1985, Kokubun1987, Christen1983}, and friction \cite{Toledo-Crow1992, Betzig1992, Karrai1995, Grober2000, Wei2000, Leuschner2001, Karrai2000, Pendry1997, Hoppe2005, Stipe2001}. 
Due to their high mechanical quality factor $Q$ between $10^3$ to $10^5$ they are very sensitive to
environmental changes. Sub-pN forces dragging on their prongs can easily be detected. 
The dragging forces change the oscillating properties of the tuning fork through a change in the spring 
constant.
This gives rise to a shift in the resonant frequency and a 
corresponding change in the amplitude of the vibrating tuning fork. 
This change in the vibration amplitude changes the current through the tuning fork, which is generated through the piezoelectric effect.
Thus, by measuring the changes of the electric signal, which can be measured very accurately, one can diagnose the prevailing forces.

This sensitivity has made quartz tuning forks become popular 
in two fields of research: temperature measurement of liquid He, where the resonance frequency is a measure for the temperature of the surrounding liquid \cite{Blaauwgeers2007, Skyba2010, Pentii2008} and in 
scanning-probe microscopy, where the forces between tip and sample are used to image surfaces with atomic resolution \cite{Toledo-Crow1992, Betzig1992, Karrai1995, Giessibl2003}. 
More and more low-temperature scanning-probe microscopes are used nowadays \cite{Rychen2000, Wintjes2010, Saitoh2008} where the height of the tip is controlled via the shear forces of a vibrating tip attached to a tuning fork. 
An accurate control of the resonance response of the tuning fork is therefore highly desired because the performance of the height control critically depends on the resonance properties of the tuning fork.
These properties are thereby described by the quality factor $Q$ of the resonance. It is defined as $2 \pi$ times the mean stored energy divided by the work per cycle or equivalently by $Q=\omega_0/\Delta \omega$, where $\omega_0$ is the resonance frequency and $\Delta \omega$ the width of the resonance. 
A high quality factor $Q$ normally results in an undesireably slow response of the system\cite{Albrecht1991} since the response time $\tau$ of
the system is related to the quality factor by $\tau={2Q}/{\omega_0}$.
Controlling the quality factor of the resonance provides a solution to this problem. It has already been shown \cite{Albrecht1991,Tamayo2000, Tamayo2001, Rodriguez2003, Morville2005, Jahng2007} that quality factor control is possible. 
Yet, up to now, it has only been applied to measurements in liquids, in ultrahigh vacuum, or at liquid He temperatures, but it has not been used to measure temperature-dependent system dynamics.

Here, we present the dynamic control of the quality factor of the resonance of a 
tuning fork over a large range of temperatures. We are able to adjust the quality factor of the resonance
to a desired value and keep it constant while changing the temperature, which is vital 
for applications purpose in near-field scanning optical microscopy.

The outline of the paper is as follows: a brief introduction into the oscillator tuning-fork system and 
its temperature dependence is followed by a description of the quality factor controlled system and
the presentation of the data showing that we are able to control the quality factor of the resonance over a wide range of temperatures. 
A short conclusion with prospects for possible experiments closes the paper.

\section{The tuning fork system}

\begin{figure}
\includegraphics[]{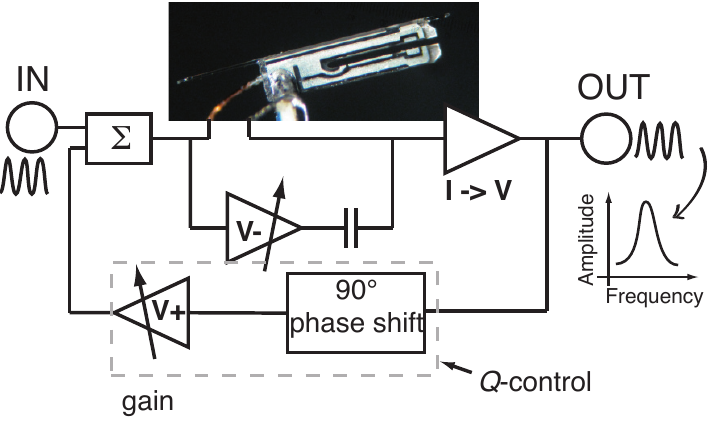}
\caption{\label{fig:fig1} Schematic of the electric circuit for the tuning-fork excitation. The parallel amplifier amd capacitance are for compensation of the parasitic capacitance of the tuning fork.
The dashed box shows the part responsible for the quality-factor control, consisting of a $90^ {\circ}$ phase shifter ($Q$-control) and an amplifier (gain).}
\end{figure}

The experiments were performed both under ambient conditions as well as in a cryostat at a temperature
between $9\ \textrm{K}$ and $300\ \textrm{K}$ and a base pressure of less than
$1 \times 10^{-4}\ \textrm{mbar}$. 
Figure \ref{fig:fig1} shows a schematic of the electric circuit to measure the electric response of the
tuning fork. A sinusoidal signal from an external function generator (IN) drives the tuning fork at a constant 
frequency with a constant amplitude. The resulting oscillation amplitude and phase shift are detected by a lock-in
amplifier (OUT). The constant frequency of the function generator acts also as a reference for the lock-in.
The additional feedback circuit (gray dashed box) consisting of an amplifier and a phase shifter 
(e.g. a fixed phase of $90^{\circ}$) is responsible for the quality-factor control.
The signal of this feedback circuit is amplified (gain) and used to excite the tuning fork in addition to the signal
from the function generator ($\Sigma$).
The phase shifter controls thereby the time delay (phase) between the excitation and the tuning-fork 
signal, which results in a damping or amplification of the tuning-fork excitation. If the driving signal and the control signal
have a phase difference of $\pi/2$ we can cancel the natural damping of the tuning fork. 
By adjusting the gain factor $g$ and  the phase (corresponding time delay $t_0$) we can increase or decrease
the quality factor $Q_{\textrm{eff}}$\cite{Rodriguez2003, Jahng2007}:
\begin{equation}
Q_{\textrm{eff}}(g,t_0)=\frac{1}{1/Q-g\sin{(\omega_0 t_0)}}.
\end{equation}

The oscillator system itself consists of a standard quartz tuning fork with an eigenfrequency of 
$\omega_0\approx 2\pi\times10^{15}\ \textrm{Hz}$. 
One prong is $l_r= 3.9$ mm long, $t_r = 600\ \mu \textrm{m}$ thick, and $w_r= 330\ \mu \textrm{m}$ wide.
In part of the experiments an optical fiber was glued to one prong of the tuning fork with epoxy glue for the use in temperature-dependent NSOM measurements. 
This changes the thickness of the prong by approximately half of the cross-sectional area of
the fiber ($t=617\ \mu \textrm{m}$) and therefore changes the response of the tuning fork compared to its natural one. 
The elastic constant of quartz and its density are $E=8.68 \times 10^{10}\ \textrm{N}/\textrm{m}^2$ and $\rho= 2649\ \textrm{kg}/\textrm{m}^3$, respectively.\cite{Lide2000} 
The resonance frequency of the tuning fork-fiber system is calculated using: \cite{Sarid1991}
\begin{equation}
\nu_0=\sqrt{\frac{\frac{E}{4}t_r(\frac{w_r}{l_r})^3}{\alpha \rho l_r t_r w_r}}=33690\ \textrm{Hz}
\label{eq:resfreq}
\end{equation}

The measured eigenfrequency of the tuning fork-fiber system is $\nu_{\textrm{meas}}=33550\ \textrm{Hz}$.
The calculation matches the experimentally determined frequency very well given the slight difference in 
values due to the evaporated contacts, the glue, and the attached fiber. 

\begin{figure}
\includegraphics[]{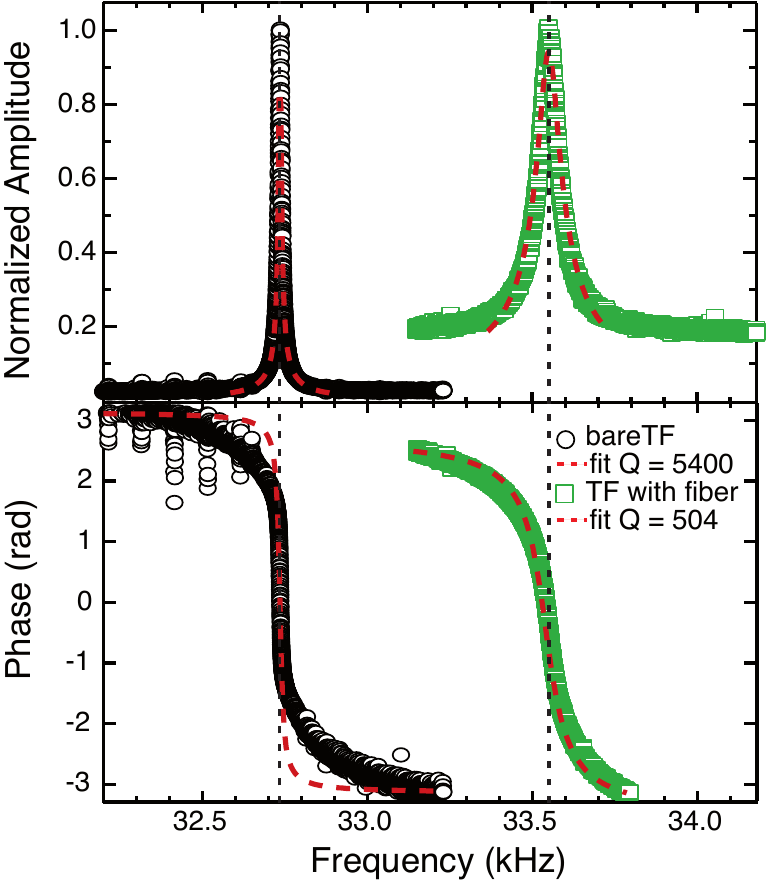}
\caption{\label{fig:fig2} (color online) Amplitude (upper panel) 
and phase (lower panel) response of a bare tuning fork (circles) and a tuning fork with a glued 
fiber (squares) at room temperature. Additionally, fitted curves (dashed lines) determining the quality factors of the two tuning forks are shown. The black dotted lines indicate the center frequency.}
\end{figure}

Figure \ref{fig:fig2} shows the amplitude (upper panel) and the phase (lower panel) response of a bare tuning fork (black circles) as well as of a tuning fork-fiber system (green squares). 
Both resonance curves were taken under ambient conditions. It is clearly visible that gluing a fiber 
to one prong of the tuning fork changes the oscillator drastically. The resonance blue-shifts 
by $1\ \textrm{kHz}$ and its amplitude is damped. 
The dashed lines in Fig. \ref{fig:fig2} show the resonances' amplitude and phase response, 
which is calculated with:

\begin{eqnarray}
A(\omega)=\frac{A_0}{m}\frac{1}{\sqrt{(\omega_0^2-\omega^2)^2+\gamma^2 \omega^2}}\\
\phi(\omega)=\arctan{(\frac{\gamma \omega}{\omega_0^2-\omega^2})},
\label{eq:amplandphase} 
\end{eqnarray}
where $\omega_0$ is the resonance frequency, $m$ is the mass of the system, $A_0$ the driving amplitude, and $\gamma$ the damping.
The model matches the experiment well and reveals quality factors $Q=\omega_0/\Delta \omega $ of
$5400$ for the bare tuning fork and $500$ for the tuning fork-fiber system, respectively. 

\begin{figure}
\includegraphics[]{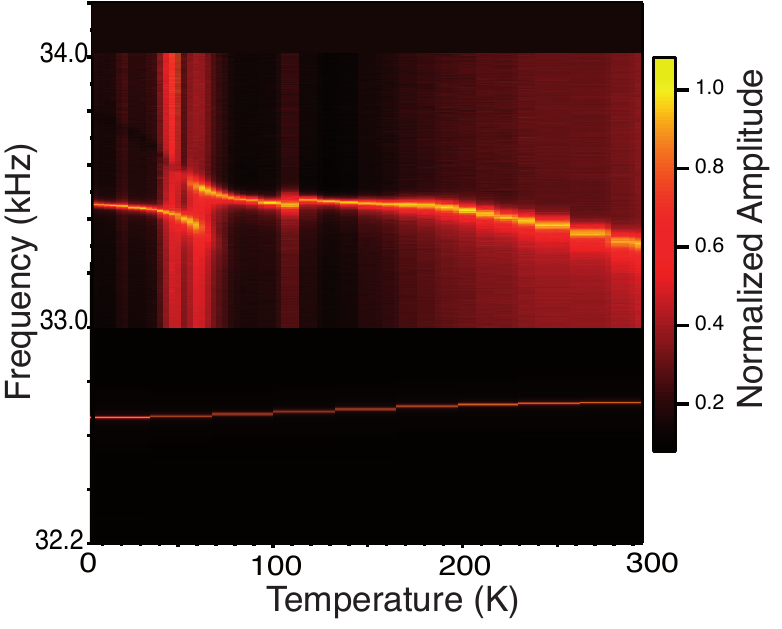}
\caption{\label{fig:fig3} (color online) Measurement of the amplitude of the oscillation of a bare tuning
fork (lower curve) and of a tuning fork with an attached fiber (upper curve) versus frequency and temperature. 
The change of the resonance frequency and width are significant in the second case.}
\end{figure}

In the next step we determined the temperature dependence of the resonance response of the bare tuning
fork and the tuning-fork-fiber system. We transferred the oscillator system into a small cryophysics
cryostat with a model 22CTI cryodyne closed-cycle helium cooler, which allowed us to measure at temperatures between 9 and 300 K.
Figure \ref{fig:fig3} shows the acquired data in a color-scale representation where the colors 
represent the amplitude of the resonance measured versus the frequency for each temperature.
The temperature-dependent response of the bare tuning fork (lower curve) shows two major features.
First of all only a small shift of the resonance frequency is measured throughout the whole temperature regime. Down to a temperature of about 
$70\ \textrm{K}$ this frequency shift is well described by:
\begin{equation}
\omega=\omega_0 [1-\alpha\cdot (T-T_0)^2],
\label{eq:tempdep}
\end{equation}
with $\alpha=3 \times 10^{-8}\ \textrm{K}^{-2}$ and $T_0=300\ \textrm{K}$. The value for $\alpha$ corresponds
thereby with the value given by the manufacturer \cite{microcrystal} and is inherent to the quartz crystal and to the 
orientation in which it is cut.
Below $77\ \textrm{K}$ the resonance shift is flattened, which is mainly due to the change in the elastic 
constants of the quartz \cite{Fukuhara1997}.    
Second, the resonance stays very sharp throughout the whole temperature range and has a quality factor
of $Q\approx 14000$ (see also Fig. \ref{fig:fig4}). 
We can assume that the measured temperature-dependent response of the bare tuning fork is universal
due to the manufacturer's design.
The effect of the vacuum in the cryostat on the resonance response can be neglected in our
experiments because it leads to a relative shift of the resonance in the order of $10^{-4}$.

In contrast to the bare tuning fork, the temperature dependence of the resonance response of 
the combined tuning fork-fiber system is more 
complex (Fig. \ref{fig:fig3}, upper curve). This is caused by the glue and the attached fiber, which 
lead to a lower quality factor compared to the bare tuning fork. Overall,
a blue-shift of the resonance frequency can be observed, which corresponds well with an
increased stiffness of the system. 
More evident is that the width of the resonance decreases with decreasing temperature. This means that at low 
temperatures the oscillator system has a higher quality factor than at room temperature. 
Another feature that is apparent in the resonance response is at $T \approx 60\ \textrm{K}$. One can clearly
observe a mode splitting and anti-crossing behavior of the resonance where the two branches interchange oscillator strength. 
While its origin is currently unclear, it depends strongly on the amount of glue used, the mass of the fiber attached, and how the 
fiber is attached to one prong of the tuning fork, thus it will be a case-by-case response.
A detailed analysis of the response is beyond the scope of the present paper.

\section{Controlling the quality factor}
\begin{figure}
\includegraphics[]{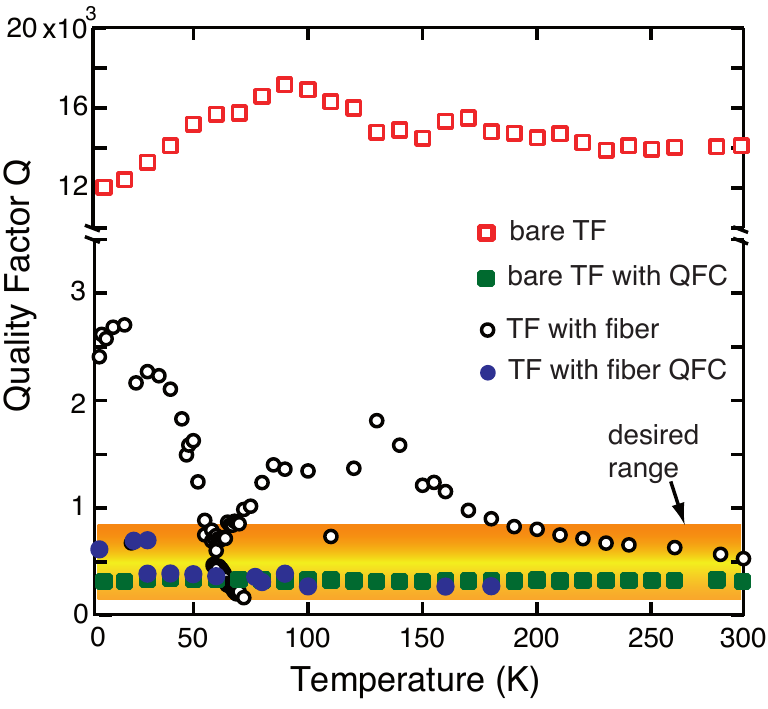}
\caption{\label{fig:fig4} (color online) Measured quality factor versus temperature for a bare tuning fork (open squares) and a tuning fork with a glued fiber (open circles). The shaded region 
indicates the desired range of quality factors where shear-force measurements are possible. Applying
quality-factor control results in a near-constant quality factor for both systems (filled symbols).}
\end{figure}

It is evident from these measurements that in order to perform shear-force controlled scanning-probe 
measurements, we need to control the quality factor of the resonance and keep it stable in the complete
temperature range. 
As a rule of thumb the quality factor of a tuning-fork system used in scanning probe microscopy (SPM) should in general lay between 300 and 1000.
In this range the feedback system for the shear-force measurements is found to be well balanced, allowing for 
stable measurements.
From our data shown in Fig. \ref{fig:fig3} we can extract the corresponding quality factors for our 
systems from the resonance width and the resonance frequency of the tuning-fork response. 
Throughout the whole temperature range these values are shown in Fig. \ref{fig:fig4} as open symbols,
squares for the bare tuning fork and circles for the tuning-fork-fiber system. 
It is seen that the bare tuning fork has quality factors exceeding $Q=12000$. More striking is the
dependence for the tuning-fork-fiber system. It is obvious that in the high-temperature region the quality 
factor is in the desired range ($Q\approx 500$). Yet, by decreasing temperature the quality factor increases by a factor of $\sim 5$ 
to $Q=2300$ at $9\ \textrm{K}$. Thereby the change in quality factor is not monotonous but changes
also due to the observed mode anti-crossing.  

The filled symbols in Fig. \ref{fig:fig4} display our results after applying our electronic phase compensation, 
\textit{i.e.}, our quality factor control. We succeeded in reducing the quality factor of both the bare tuning fork
and the tuning-fork-fiber system and kept it nearly constant throughout the whole temperature range. 

\begin{figure}
\includegraphics[]{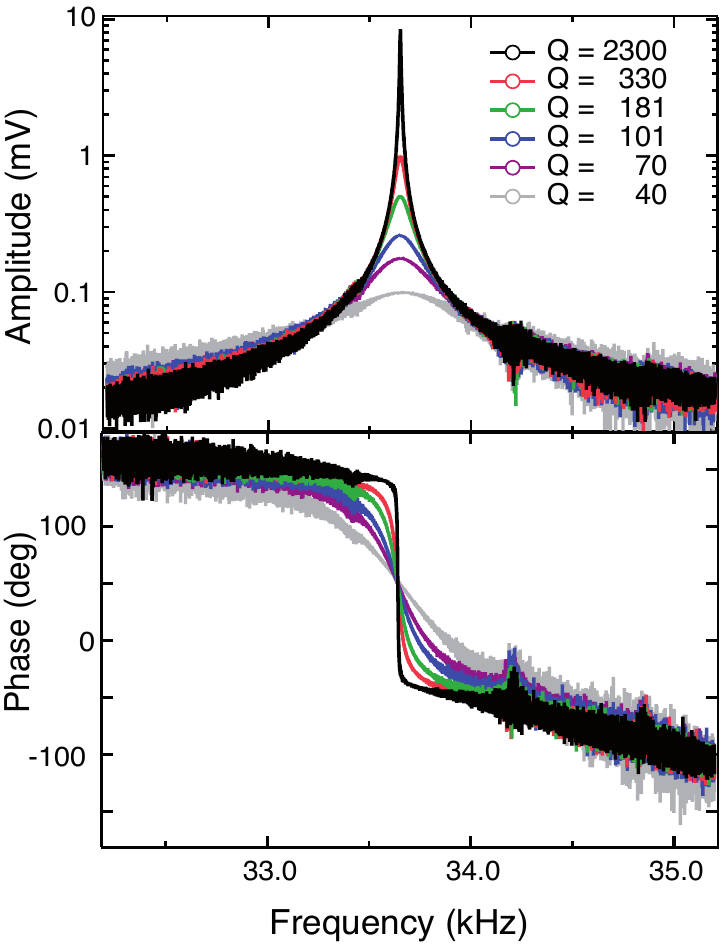}
\caption{\label{fig:fig5} (color online) Quality-factor control measurements of the resonance at $T=9\ \textrm{K}$ for a tuning fork with an attached fiber.
(Upper panel) Amplitude response of the tuning fork. The quality factor of the resonance was tuned between $Q=40$ and $2300$. 
(Lower panel) Simultaneously recorded phase of the oscillator.}
\end{figure}

To test our system's dynamic performance we continuously changed the gain settings for the quality-factor
control. Figure \ref{fig:fig5} shows the resonance response of a tuning fork-fiber system at a temperature of $9\ \textrm{K}$.
We recorded the amplitude (upper panel) as well as phase (lower panel) versus the frequency. The starting
value of $Q=2300$ (compare also Fig. \ref{fig:fig4}) is achieved without any quality factor control. As is evident 
from the graph, we can change the quality factor of the tuning fork smoothly down to $Q=40$ by changing the
gain in the phase shifter of the quality-factor-control unit (Fig. \ref{fig:fig1}, gray box). The good performance of our quality factor control is 
also apparent in the recorded phase, a smooth transition from high- to low-$Q$ is visible.

The only restricting factor still prominent in the measurements is the noise level.  
For lower quality factors the signal-to-noise level is the limiting parameter to resolve the resonance of the tuning fork clearly. 
If the main noise source is the electronics that were used, this restriction can be reduced significantly. Applying an electronic 
compensation scheme where the reference signal and the signal from the tuning fork are subtracted, 
we achieved to reduce the noise level by at least a factor of 10 (data not shown here), thereby increasing  
the performance of the system for the entire range of quality factors. 

\begin{figure}
\includegraphics[]{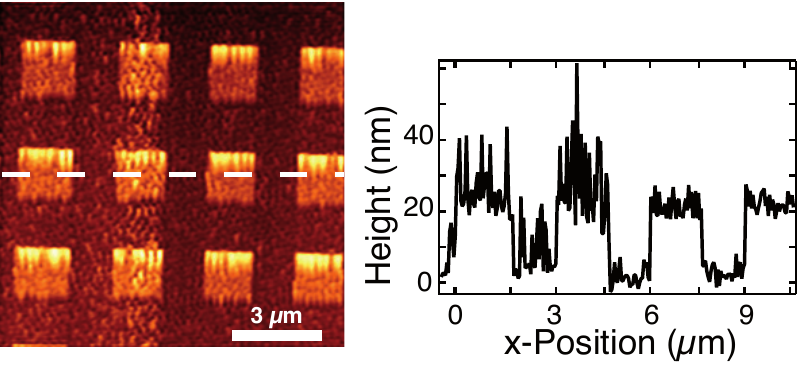}
\caption{\label{fig:fig6} (color online) Shear-force topography scan of a test grating. For this scan an adjusted quality factor of $QÊ\approx 350$ was used. Between the left and right parts of the image the gain in the height feedback control was changed. (right panel) The cross section was taken along the dashed line in the topography image.}
\end{figure}

With the uncontrolled high value for $Q$ we could not achieve any shear-force controlled measurement of 
a topography. Yet, by changing the quality factor and keep it constant ($Q \approx 350$) we are now able
to measure the topography of a standard test grating \cite{nmdt} with our tuning fork-fiber system, as
shown in Fig. \ref{fig:fig6} (left panel). 
The displayed squares have a height of $20\ \textrm{nm}$, a pitch size of $1.5\ \mu \textrm{m}$, and a periodicity of 
$3\ \mu \textrm{m}$. There is a difference between the left and right part of the recorded image visible. 
The left part has more noise along the fast scan direction of the microscope. 
Changing the gain of the height feedback halfway during the scan led to reduced noise and a higher contrast in the topography. 
This is visible in more detail in the cross sectional scan (Fig. \ref{fig:fig6}, right panel) along the dashed line
of the topography image. The height of the pitches deduced from this scan matches the height given by 
the manufacturer and shows the good perfomance of the shear-force quality-factor controlled scanning system.

\section{Conclusions and Outlook}

We presented our results on the dynamic control of the quality factor of a quartz tuning fork in a large
temperature range. We showed that we are able to compensate for the temperature-dependent change
in the resonance response of a tuning fork. Moreover, we applied our compensation to a tuning fork-fiber 
system and achieved to keep the quality factor in a regime where shear-force measurements are possible
throughout the whole temperature range. We are also able to smoothly tune the quality factor at each 
temperature to a certain value.
To be able to measure topography with the same quality factor of the tuning fork is a major step to
a more efficient and straightforward use of such scanning-probe microscopy systems and allows their use
over the complete temperature range as well as in all sorts of environments.

As a result of our experiments, dynamic temperature-dependent near-field optical measurements have become possible. The method can be used, \textit{e.g.}, to measure locally resolved structural phase transitions, such as domain formation, by optical means with varying temperature and thus deduce the optical properties of sub-wavelength structures near a transition point. Furthermore, the temperature dependence of light propagation in nanophotonic objects can be studied.

\section{Acknowledgments}
 We would like to
thank Frans Segerink, Allard Mosk, Herman Offerhaus, Cock Harteveld, and
Dick Veldhuis for discussions and help with setting up equipment. 
PP acknowledges financial support by FOM Projectruimte. This work is also part of the research program of FOM, which is financially supported by NWO.

\end{document}